\documentclass[12pt]{article}
\usepackage[utf8]{inputenc}
\usepackage{graphicx}
\usepackage{listings}
\usepackage{hyperref}
\usepackage{xcolor}
\usepackage{geometry}
\usepackage{csquotes}
\title{JavaSith: A Client-Side Framework for Analyzing Potentially Malicious Extensions in Browsers, VS Code, and NPM Packages}
\author{Avihay Cohen}
\date{May 2025}

\begin{document}

\maketitle

\begin{abstract}
Modern software supply chains face an increasing threat from malicious code hidden in trusted components such as browser extensions, IDE extensions, and open-source packages. This paper introduces \textit{JavaSith}, a novel client-side framework for analyzing potentially malicious extensions in web browsers, Visual Studio Code (VSCode), and Node's NPM packages. JavaSith combines a runtime sandbox that emulates browser/Node.js extension APIs (with a ``time machine'' to accelerate time-based triggers) with static analysis and a local large language model (LLM) to assess risk from code and metadata. We present the design and architecture of JavaSith, including techniques for intercepting extension behavior over simulated time and extracting suspicious patterns. Through case studies on real-world attacks (such as a supply-chain compromise of a Chrome extension and malicious VSCode extensions installing cryptominers), we demonstrate how JavaSith can catch stealthy malicious behaviors that evade traditional detection. We evaluate the framework's effectiveness and discuss its limitations and future enhancements. JavaSith's client-side approach empowers end-users/organizations to vet extensions and packages before trustingly integrating them into their environments.
\end{abstract}

\section{Introduction}
Malicious code injection in trusted software components has emerged as a major cybersecurity threat. In particular, browser extensions, IDE extensions, and open-source packages have been weaponized as vectors to infiltrate systems via supply chain attacks. Recent incidents underscore the scale of the problem: in late 2024, a compromised Chrome extension update injected data-stealing code and affected over 400,000 users\cite{darktrace2024}, as part of a broader campaign compromising 35 extensions and impacting 2.6 million users\cite{hunters}. Likewise, in 2025, researchers discovered ten malicious VS Code extensions (masquerading as popular tools with hundreds of thousands of installs) that surreptitiously deployed cryptocurrency miners on developers' machines\cite{broadcom2025}. These examples highlight how threat actors leverage the trust and reach of extensions and packages to compromise large user bases. Alarmingly, such malicious extensions often evade official store security checks – for example, the malicious Chrome extension update in the Cyberhaven incident passed Google's Web Store review process.

Traditional defenses struggle against these emerging threats. Signature-based antivirus rarely flags novel extension malware, and permission-based vetting is insufficient when attackers find creative ways to abuse allowed capabilities. Supply chain attacks via extensions and packages have surged by 431\% in recent years\cite{insurancebusinessmag}, indicating that adversaries are actively adapting to exploit blind spots. The need for better tooling to analyze and detect malicious behavior in extensions before damage occurs is evident.

In this paper, we introduce \textbf{JavaSith}\cite{javasith}, a client-side framework that empowers researches/organizations to scrutinize browser extensions\cite{javasith}, VSCode extensions\cite{javasith_ide}, and NPM packages\cite{javasith_code} for malicious intent. Unlike server-side scanning services, JavaSith runs locally in the user's browser (for privacy and immediacy), integrating multiple analysis techniques:
(1) a dynamic \textit{sandbox} entirely in the browser execution context that safely executes extension code in an instrumented environment, emulating browser extensions api or Node.js APIs so that all invoked calls and behaviors can be observed;
(2) a \textit{time machine} module that can fast-forward or simulate time to trigger time-dependent logic (e.g., delayed or periodic malicious payloads);
(3) a \textit{code parser} module that parsed javascript code and converts it to AST for further insights (e.g function names , variables, invoked APIs ...)
(4) a static analysis pipeline incorporating tools like Retire.js\cite{retirejs2024} (to detect known vulnerable libraries) and custom detectors for obfuscation or suspicious API usage;
(5) a lightweight on-device \textit{WebLLM}\cite{webllm2023} (Web-based Large Language Model) that assesses risk by analyzing the extension's source code, metadata, and even privacy policy text to identify potential red flags in plain language;
(6) a user-friendly web-based GUI that presents the comprehensive analysis results, including raw logs, flagged patterns, and LLM-generated summaries of risky behavior.

Our aim is to provide a holistic analysis that can catch complex, stealthy malicious behaviors that single techniques might miss. For instance, dynamic monitoring can reveal if an extension tries to exfiltrate data or fetch external payloads at runtime, while static inspection can identify hidden backdoors or suspicious strings (like hardcoded webhook URLs or encrypted payloads). The inclusion of an LLM-based analysis offers a higher-level interpretation of the findings, bridging the gap between raw technical signals and an understandable risk assessment for the user.

We validate JavaSith through case studies of known malicious extensions and packages. We show that JavaSith would have flagged the malicious Chrome extension involved in the Cyberhaven breach by detecting its cookie-stealing code and unusual network behavior, even if it remained dormant until a specific time window. In the case of VSCode cryptominer extensions, our sandbox observes the download of a PowerShell script and subsequent installation of a hidden cryptominer, actions that we surface as severe warnings to the user. We also demonstrate JavaSith on an example malicious NPM package that uses obfuscated install scripts, illustrating how combining static and dynamic analysis can uncover the hidden payload.

In summary, the contributions of this work include:
\begin{itemize}
    \item \textbf{JavaSith Framework}: We design and implement a novel client-side framework to analyze browser extensions, VSCode extensions, and NPM packages for malicious behavior. JavaSith combines runtime sandboxing, time acceleration, static analysis, and LLM-based code understanding in a unified tool that is completely client based.
    \item \textbf{Runtime Sandbox with Time/Events Simulation}: We introduce a sandbox that emulates the targeted runtime (browser or Node) and intercepts extension API calls, DOM manipulations, and network requests. A built-in ``time machine'' can accelerate timers and simulate future dates to trigger logic bombs or delayed payloads that would otherwise require long real-time waiting. the sandbox also simulates different set of events like page navigation to sensitive domains like login.microsoftonline.com or facebook.com and then monitor the extension code for malicious bahviour like reading cookies / injecting javascript / ...
    \item \textbf{Integrated Static + LLM Analysis}: JavaSith integrates static vulnerability scanning (using tools like Retire.js for known vulnerable libraries\cite{retirejs2024} and custom pattern matching for suspicious code constructs) with a WebLLM-based analysis that summarizes and highlights risky behavior from the extension's source code and metadata. By running the LLM locally in the browser via WebGPU, we ensure privacy while leveraging advanced reasoning over code.
    \item \textbf{Evaluation on Real Attacks}: Through detailed case studies (Chrome extension supply-chain attack, malicious VSCode cryptominer extensions, and others), we demonstrate that JavaSith can detect malicious behaviors (cookie/session theft, unauthorized network beaconing, cryptomining installation, credential exfiltration via webhooks, etc.) that might evade detection by conventional means. We measure JavaSith's effectiveness in triggering hidden behaviors and its performance overhead on sample extensions.
    \item \textbf{Tool for the Community}: We provide a browser-based graphical interface that allows users to navigate analysis results, including raw execution traces, flagged security issues, and explanatory risk reports generated by the LLM. This interface lowers the barrier for developers and security analysts to vet third-party extensions and packages, complementing official marketplace security measures.
\end{itemize}

\section{Background and Motivation}
Browser extensions and plugin-based software ecosystems greatly expand functionality but also inherently expand the attack surface. Here we outline how browser and VSCode extensions operate, the current security measures in place, and the ways attackers have exploited these platforms, motivating the need for JavaSith.

\subsection{Extension Architecture and Security Model}
\textbf{Browser Extensions:} Modern browsers (Chrome, Firefox, etc.) allow extensions to run privileged scripts with capabilities beyond normal web pages. Extensions typically consist of a background script (with persistent privileges), content scripts injected into web pages, and optional UI components. They declare permissions (such as access to network, cookies, tabs, or specific domains) in a manifest. While browsers enforce some isolations (content scripts are isolated from page scripts), a granted permission like `cookies` or `<all\_urls>` can let an extension read sensitive user data or modify web content. Security of extensions largely relies on user trust and store vetting, as once installed, an extension's code executes with the user's privilege and can perform any allowed actions. Notably, extension updates are automatic, meaning an attacker who gains control of an extension's distribution (e.g., by hijacking a developer account) can push malicious code to all users silently.

\textbf{VS Code Extensions:} VSCode extensions run inside an extension host process with the same privileges as the user running VSCode\cite{vscode_security2024}. This means a VSCode extension can read/write any file on the machine, make network requests, and run external processes. VSCode has introduced a few safeguards (for instance, prompting the user to trust the publisher of an extension on first install\cite{vscode_security2024}, and a Workspace Trust feature that can restrict automatic code execution in untrusted projects). The VS Code Marketplace also performs automated malware scanning and even dynamic analysis in a sandboxed environment for submitted extensions\cite{vscode_security2024}. However, as recent incidents show, malicious extensions still slip through these defenses. Because VSCode extensions are essentially Node.js programs, attackers can leverage Node APIs (child processes, filesystem, etc.) within an extension. This broad capability makes it challenging to sandbox extensions without breaking functionality.

\textbf{NPM Packages:} The NPM ecosystem for Node.js allows developers to pull in millions of packages. Malicious actors have published trojan packages (often via typosquatting or by compromising maintainer accounts) that execute harmful code during installation or runtime. NPM packages can include install scripts that run upon `npm install`, giving immediate execution of code on the developer's machine. There is minimal sandboxing—if a developer installs a malicious package, it can, for example, read environment variables (to steal secrets), modify files, or download and execute binaries. While NPM and services like GitHub have security advisories and automated scans for known malware, novel attacks can bypass signature checks through obfuscation or multi-stage payloads.

\subsection{Threats and Notable Incidents}
Supply chain attacks leveraging these extension ecosystems have grown in frequency and sophistication. Attackers exploit the implicit trust users place in extensions and the difficulty of manual review at scale.

One prominent example is the Cyberhaven Chrome extension incident in December 2024. Attackers phished a Chrome Web Store developer credential, allowing them to push a malicious update to a legitimate extension\cite{darktrace2024}. The malicious version added hidden scripts (`worker.js` and `content.js`) to exfiltrate cookies, session tokens, and user data, particularly targeting Facebook accounts\cite{darktrace2024}. The timing was deliberate—released during a holiday period when oversight was low. This single extension compromise affected over 400,000 users and formed part of a larger campaign across dozens of extensions\cite{hunters}. Because the malicious code was blended into an otherwise normal extension and even passed store review, victims’ security tools did not flag it. Traditional network security only noticed anomalies when the extension began beaconing out to a suspicious domain disguised to look legitimate (e.g., ``cyberhavenext.pro'').

In the VS Code ecosystem, a recent campaign in April 2025 saw multiple malicious extensions uploaded to the Marketplace under guises like “Prettier” or “Golang Compiler.” These extensions included code that, upon activation, fetched a remote PowerShell script from an attacker-controlled server and executed it to install a Monero cryptominer\cite{broadcom2025}. Cleverly, after running the malicious payload, the extension would install the real legitimate extension it impersonated, to avoid arousing user suspicion. The PowerShell script took extensive steps to persist and hide itself: disabling Windows security services, adding scheduled tasks, and launching the miner with elevated privileges. Investigators found that the attacker's server contained hints of a parallel campaign targeting NPM packages as well, illustrating how attackers reuse tactics across ecosystems.

Another case involved VSCode extensions designed to steal information. ReversingLabs uncovered several extensions (e.g., `clipboard-helper-vscode`, `codegpt-helper`) that quietly siphoned off sensitive data via web requests to a hardcoded Discord webhook URL\cite{reversinglabs2025}. For instance, one such extension monitored the user's clipboard and, whenever text was copied, sent the clipboard content to the attacker's Discord channel. In another variant, an extension targeted credentials by reading an ``easycode.openAI ApiKey`` setting (which might store an API key for a code assistant) and exfiltrated it via Discord as well. Discord webhooks are a common exfiltration mechanism for malware, since traffic to Discord may appear benign and can bypass some firewall rules. The presence of a Discord webhook URL in extension code is therefore a strong indicator of malicious intent.

The NPM package ecosystem has similarly seen sophisticated malware. For example, a malicious package `os-info-checker-es6` was discovered in 2025 that used invisible Unicode characters in its source to hide malicious logic, and it stored its command-and-control URL in a Google Calendar invite as an obfuscation tactic\cite{thehackernews}. Initially, this package was uploaded as a benign utility, but a later update introduced platform-specific binaries and obfuscated install scripts that pulled in a final payload from the hidden C2 link\cite{thehackernews}. The malware would collect OS details and likely attempt to download additional components. Such tricks defeat naive static scans and show why dynamic analysis (executing the package in a safe sandbox) is invaluable.

These incidents motivate the design of JavaSith. In each case, either dynamic behavior (network calls, timed triggers, environment changes) or static indicators (suspicious strings, obfuscated code) were present that could tip off a careful analysis. Official protections (like Chrome Web Store and VS Code Marketplace scans) were bypassed or not comprehensive enough. A defense-in-depth approach on the user side, where the extension or package is vetted in a controlled environment with multiple detection layers, can significantly reduce the risk of installing such malicious components.

\section{Related Work}
Prior work on detecting malicious extensions and packages spans both industry tools and academic research. Browser extension security has been the focus of several static and dynamic analysis systems. Google’s Chrome Web Store employs automated scanning and manual review, but as noted, attackers increasingly design malware to evade these. Third-party tools like \textit{CRXcavator} have emerged to evaluate Chrome extensions by analyzing their code and metadata for risk factors (e.g., dangerous permissions, suspicious API usage). Similarly, in the Node ecosystem, package auditing tools (e.g., \textit{npm audit}) and services like Socket.dev attempt to flag malicious or vulnerable packages by scanning for known indicators such as obfuscated code or unusually broad permissions.

Academic research has explored detecting malicious browser extensions using both static and dynamic techniques. Some approaches analyze the differences between extension versions (updates) to spot injected malicious code, while others use information flow analysis to check for suspicious data exfiltration. There have also been machine learning methods using extension metadata, descriptions, and permission sets to classify likely malware. However, these methods often suffer from high false positives or can be evaded by clever attackers.

For VS Code extensions, the threat is relatively newer. The VS Code team’s introduction of publisher trust verification and sandbox tests\cite{vscode_security2024} is a direct response to the possibility of malicious extensions. Yet, as the April 2025 cryptominer case shows, dynamic detection can miss payloads that trigger only under certain conditions (e.g., post-install or requiring user action). ReversingLabs’ research\cite{reversinglabs2025}, as well as Check Point’s blog posts, have shed light on early malicious VSCode extensions, primarily by manually analyzing suspicious extensions found in the wild.

Our work is also related to general malware sandboxes and dynamic analysis environments commonly utilized for analyzing executables and mobile applications. Existing tools, such as Cuckoo Sandbox and various mobile app sandboxes, execute code within isolated environments to detect malicious behaviors. JavaSith adopts a similar methodology but extends it specifically to the domain of browser extensions and software packages, addressing unique challenges such as accurately simulating browser or Visual Studio Code (VSCode) environments within a browser-based analysis setting. A distinguishing aspect of our approach is the incorporation of recent advancements in machine learning, particularly on-device Large Language Models (LLMs), to facilitate dynamic analysis. While machine learning-based code analysis is actively explored in research—with previous works leveraging neural networks for vulnerability detection or malware classification—our framework introduces a novel application by employing a local LLM at runtime. This enables the interpretation of an extension's intended functionality and associated risks dynamically. Moreover, the LLM module evaluates the extension's privacy policy and correlates its stated intentions against observed behaviors. Specifically, if an extension claims in its privacy policy not to transmit sensitive user data but the sandbox detects sensitive data exfiltration, the LLM can explicitly highlight this contradiction, thus providing a strong indicator of potentially malicious intent.

In summary, JavaSith differentiates itself by unifying multiple analysis techniques tailored to the extension/package context and running entirely client-side. This avoids the privacy concerns of uploading code to cloud scanners and allows integration directly into user workflows (e.g., a developer could run JavaSith on a new VSCode extension before enabling it, or on a library before adding it to a project). Our contributions complement existing store-level protections by focusing on the end-user’s ability to independently verify code trustworthiness.

\section{Design and Architecture of JavaSith}
\begin{figure}[h]
\centering
\includegraphics[width=0.8\textwidth]{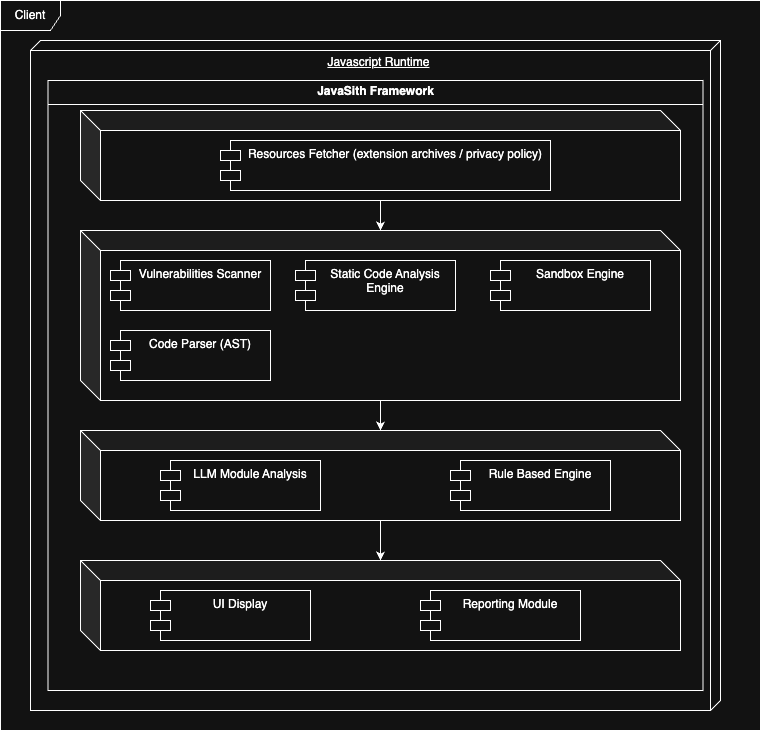}
\caption{High-level architecture of JavaSith. The framework ingests extension or package code, then performs static analysis, dynamic execution in a sandbox (with time simulation), and LLM-based evaluation. A GUI allows the user to review detailed findings.}
\label{fig:architecture}
\end{figure}

JavaSith comprises several coordinated components that together perform comprehensive analysis of an extension or package. Figure~\ref{fig:architecture} provides an overview of the architecture. The core components include: (i) the Emulated Runtime Sandbox, (ii) the Time Machine, (iii) the Static Analysis Engine, (iv) the WebLLM Risk Analyzer, and (v) the Results Dashboard GUI. We describe each in detail below.

\subsection{Emulated Runtime Sandbox}
\begin{figure}[h]
\centering
\includegraphics[width=0.8\textwidth]{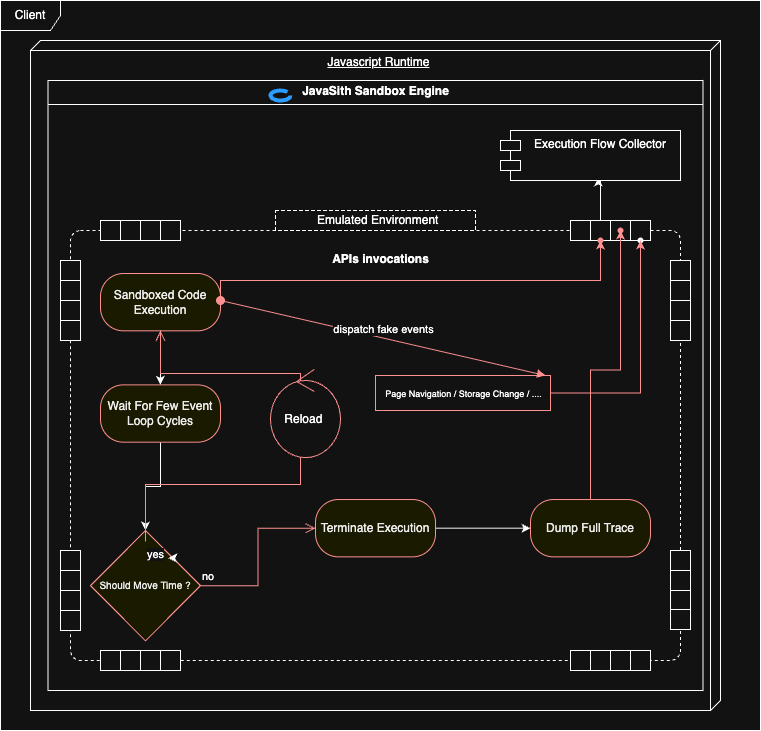}
\caption{High-level architecture of JavaSith Novel Sandbox.}
\label{fig:sandbox_architecture}
\end{figure}

At the heart of JavaSith is a sandbox environment that can execute the code of the extension or package in question, but in an instrumented manner that logs all interesting actions. Because we target three contexts (browser, VSCode, Node package), the sandbox must flexibly emulate each environment:
\begin{itemize}
    \item For \textbf{browser extensions}, JavaSith emulates the Chrome extension runtime. We provide implementations of the Chrome extension API (such as \texttt{chrome.runtime}, \texttt{chrome.storage}, \texttt{chrome.tabs}, etc.) in a controlled isolated context. Content scripts from the extension are loaded into a separated isolated execution context (using, for example, SES compartment \cite{js_ses}), and background scripts are run in isolated context that mimics service worker context. We intercept calls like \texttt{chrome.webRequest} or any network fetches, file access, or cookie access. For each API call, the sandbox logs the call details (function name, arguments) and can optionally block the real effect (e.g., instead of actually deleting a file or exfiltrating data, we record the attempt).
    \item For \textbf{VSCode extensions}, we create a fake VSCode extension host environment. This involves providing a stub implementation of the VSCode extension API (the \texttt{vscode} module) so that when the extension activates, it doesn’t throw errors for missing APIs. We run the extension’s activation function in a emulated \texttt{vm} context (sandbox) so that we can monitor its behavior. Any calls to Node built-ins (like \texttt{fs}, \texttt{child\_process}) are tracked. The sandbox here closely mirrors an actual VSCode extension host but ensures no harmful actions escape (e.g., file writes are redirected to a temporary sandbox filesystem, network calls are intercepted and logged or stubbed out, etc.).
    \item For \textbf{NPM packages}, since these are plain Node.js modules, the sandbox essentially runs \texttt{npm install} and then imports the package in an isolated environment. We pay special attention to the install/postinstall scripts: JavaSith will execute those in the sandbox and monitor their behavior. Similarly, if the package exports functions, we can optionally exercise some of them if known (though automated function exercising is limited; we focus on installation and initialization behaviors by default).
\end{itemize}

To implement the sandbox, we leverage SES Compartments \cite{js_ses}. JavaSith can override global functions like \texttt{setTimeout}, \texttt{XMLHttpRequest}/\texttt{fetch} (in the browser context), and \texttt{child\_process.exec} (in Node) to hook into their behavior. For instance, network requests are routed through a proxy method that logs request URLs and data. File system calls are redirected to a in memory sandbox directory.

The sandbox design emphasizes both monitoring and safety. Because our sandbox runs within a browser context, it benefits from the same robust isolation provided by the browser's native sandboxing mechanisms. This ensures that if the code attempts something destructive (such as wiping the file system or launching ransomware), it cannot affect the host system. In practice, extensions typically won't attempt such overt destructive actions—since these would immediately expose their malicious intent—but they might attempt subtle actions like quietly exfiltrating data, which we can safely observe and monitor within this isolated environment.

All events recorded by the sandbox are timestamped and categorized (e.g., “Network request to X,” “Read file Y,” “Called chrome.tabs.query,” etc.). These events form a timeline of the extension’s behavior during analysis.

\subsection{Time Machine Module}
Malicious extensions may deliberately delay their payloads or trigger them only under specific timing conditions (for example, only after being installed for a week, or on a certain date, or gradually over time). To catch such logic, JavaSith includes a Time Machine module that can manipulate the perceived time within the sandbox.

The Time Machine hooks timer functions and date/time APIs. For instance, \texttt{setTimeout} and \texttt{setInterval} calls are intercepted so that the analysis can fast-forward their execution. We replace the standard timers with our own that can simulate the passage of time. Similarly, reads of the system clock (e.g., \texttt{Date.now()}) can be overridden to return a simulated time if needed.

In practice, the user (or the system automatically) can advance the sandbox clock or trigger scheduled events. For example, if the extension sets a timeout to execute after 24 hours, our system will not literally wait 24 hours; instead, it will fast-forward the clock and immediately execute the timeout callback. Listing~\ref{lst:timemachine} shows a simplified pseudocode of how this works.

\begin{lstlisting}[caption=Simplified Time Machine Hook for Timers,label=lst:timemachine]
const originalSetTimeout = global.setTimeout;
global.setTimeout = function(fn, delay, ...args) {
    log(`Timer set for ${delay}ms -> fast-forwarded execution.`);
    // Immediately (or after a shortened delay) execute the callback:
    return originalSetTimeout(fn, Math.min(delay, MAX_WAIT_MS), ...args);
};
\end{lstlisting}

We also handle recurring timers (\texttt{setInterval}) similarly, and allow manual time jumps (e.g., simulate that one week has passed) which can affect code using \texttt{Date} or \texttt{new Date()}.

A special case of time-triggered logic is where malicious code checks the current date against a threshold (a logic bomb). For instance, “only execute payload if current date > June 1, 2025”. The Time Machine can detect when code is checking the date and optionally manipulate the returned date. However, identifying such checks reliably via dynamic means can be tricky unless the code executes that branch; this is where static analysis (scanning for date references) can complement, or a user can instruct the sandbox to set a custom current date.

By enabling time control, JavaSith can trigger behaviors that would otherwise require patience or might never occur during a short dynamic run. We found this particularly useful in the Chrome extension case study, where the extension only activated data theft when certain conditions were met (e.g., a specific site was visited during a certain time window). By simulating those conditions, including the passage of time and triggering of alarm events, we increase coverage of the extension’s potential behaviors.

\subsection{Static Analysis Engine}
While dynamic analysis is powerful, it may not execute every code path. Thus, JavaSith performs static analysis on the extension/package code both before and after dynamic execution. The static analysis engine has several facets:

\paragraph{Dependency and Vulnerability Scanning:} Using the Retire.js database, JavaSith identifies known vulnerable libraries embedded in the code\cite{retirejs2024}. For example, if an extension bundles an outdated jQuery version with a known XSS vulnerability, that might not be malicious per se, but it raises the risk profile. More pertinently, if an extension includes a library that is known to be commonly used in malware (for instance, a specific cryptominer WebAssembly module or a credential-stealing snippet), that is a red flag.

\paragraph{Heuristic Pattern Analysis:} We scan for patterns like suspicious function names or usage of dangerous APIs. For browser extensions, usage of \texttt{eval} or dynamic code generation, heavy obfuscation (e.g., a high ratio of non-alphanumeric characters or long unreadable variable names), or references to external URLs (especially if hard-coded) are noteworthy. For VSCode/Node, usage of \texttt{child\_process.exec} or launching PowerShell, or the presence of encoded payload strings (like large Base64 blobs) are flagged. For instance, in the malicious VSCode extensions, the code snippet that fetched a PowerShell script from `asdf11.xyz` would be caught by a rule looking for network calls to uncommon domains.

We also check for Discord webhook URLs (`discord.com/api/webhooks`) since, as noted, that is strongly correlated with data-stealing extensions. If found, we mark the extension as likely exfiltrating data.

\paragraph{Privacy Policy and Metadata Check:} Many extensions provide a privacy policy or description. We ingest these (if available in the extension package or store metadata) and apply NLP (and the onboard LLM) to see if the text aligns with observed behaviors. For example, an extension claiming to never collect personal data while the code clearly accesses cookies and sends data out is suspicious. Although this is not purely static code analysis, it is analysis of static metadata that can reveal inconsistencies or misrepresentations.

The static engine produces a report of findings: e.g., “Found reference to Discord webhook URL,” “Contains base64-encoded binary of 200KB (possible embedded payload),” “Uses eval on obfuscated string,” “Bundles jQuery v1.12.0 with known vulnerabilities,” etc. Each finding is cross-referenced with source file and line if possible.

These static results feed into the next component, the LLM analyzer, and also directly to the GUI for expert users to inspect.

\subsection{WebLLM-Based Risk Analyzer}
JavaSith integrates a local large language model to synthesize the findings from dynamic and static analysis and provide a human-readable assessment of the extension’s risk. This component leverages \textit{WebLLM}\cite{webllm2023}, a framework that allows running moderately large LLMs (such as a 7B-13B parameter model) directly in the browser via WebGPU acceleration. We fine-tuned or prompt-engineered the LLM for security analysis tasks, guiding it to output a structured review of the extension.

The input to the LLM includes:
\begin{itemize}
    \item A summary of static findings (e.g., “This extension accesses cookies and contacts \texttt{cyberhavenext.pro} domain”).
    \item A summary of dynamic findings (e.g., “During execution, the extension made 5 network requests to suspicious domains and attempted to install an external program.”).
    \item Possibly snippets of de-obfuscated code or formatted listings of key functions for the LLM to reason about.
    \item The extension’s manifest and description, so the LLM knows the purported purpose.
    \item The extension privacy policy , so the LLM can give a privacy risk score.
\end{itemize}

Given this context, the LLM is prompted with something like: “Analyze the provided extension behavior and code for potential malicious intent. Explain what the extension is doing and why it may be dangerous. Rate the risk level as High, Medium, or Low, with reasoning.” The model then generates text such as: \textit{“This Chrome extension appears to steal user session cookies and send them to an external server (cyberhavenext.pro) under the guise of a legitimate domain. It also strips security headers (Content-Security-Policy) from websites, which is behavior indicative of trying to enable further attacks. This behavior is highly malicious as it can lead to account takeover. Risk level: High.”} 

The local LLM approach ensures that even sensitive code (perhaps proprietary) can be analyzed without sending it to a cloud API, alleviating privacy concerns. The trade-off is that the model we can run in-browser is smaller than state-of-the-art, but we found that an open-source model fine-tuned for code analysis is effective at summarizing suspicious behaviors and pointing out dangers that a layperson might not immediately glean from raw logs.

The LLM’s output is not taken as ground truth but as an aid. It might occasionally misinterpret code; therefore, we always pair its narrative with the factual raw data from other modules. Still, it adds significant value in translating technical signals into a summary (much like an analyst’s report). Users can thus see something like “LLM Analysis: This extension likely steals your session tokens and data from certain websites” alongside the concrete evidence.

\subsection{Graphical User Interface (GUI)}
JavaSith operates within the client environment (web browser) and presents analysis results through a state-of-the-art graphical user interface (GUI). The GUI is integral for usability, as it effectively organizes analytical outputs into distinct tabs:

\begin{itemize}
\item An \textbf{Overview} tab providing a comprehensive summary of all metadata gathered from the relevant store or manifest.
\begin{figure}[h]
\centering
\includegraphics[width=0.8\textwidth]{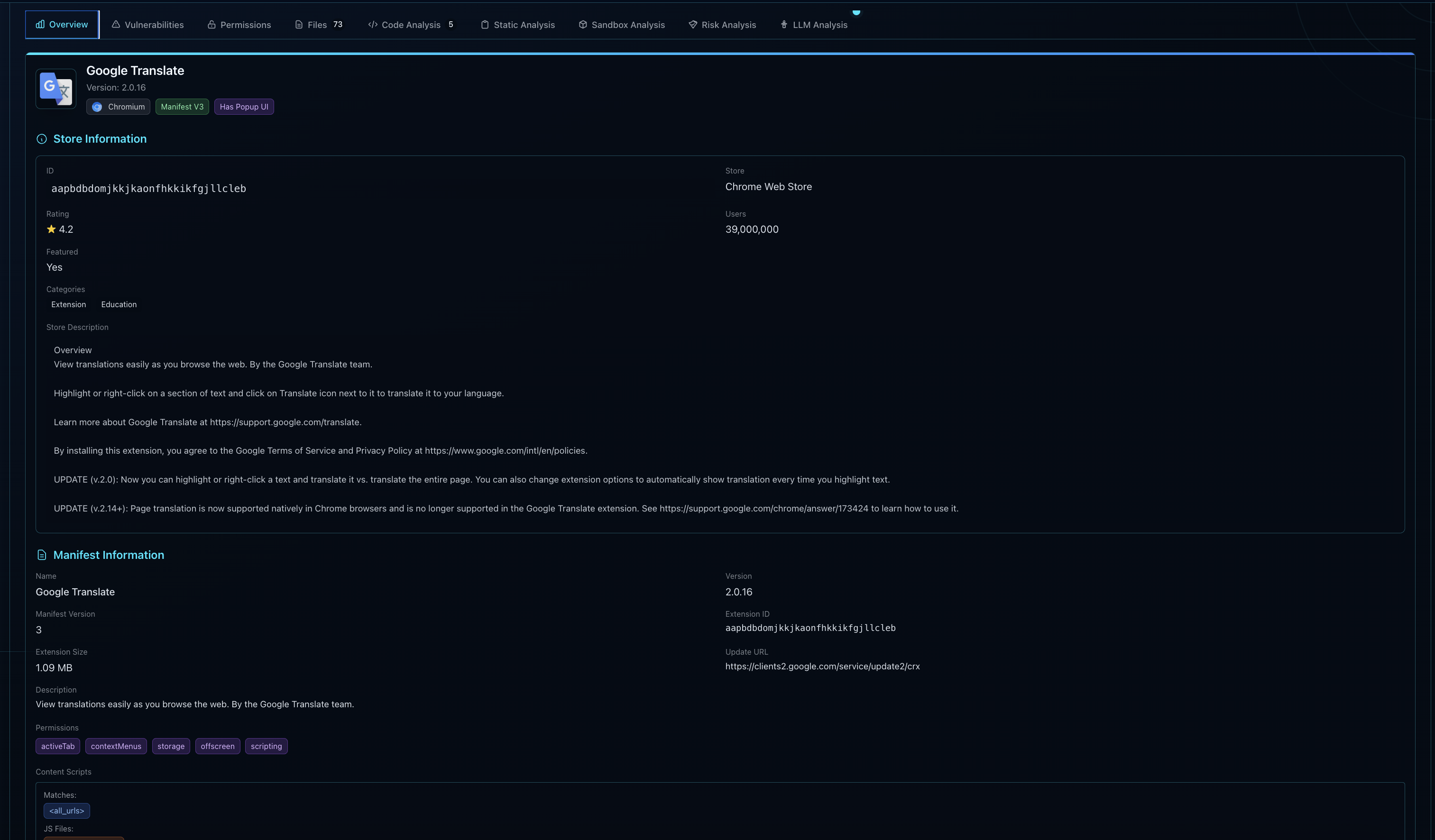}
\caption{Overview Tab}
\label{fig:overview_tab}
\end{figure}
\item A \textbf{Vulnerabilities Summary} highlighting detected vulnerabilities within the code.
\begin{figure}[h]
\centering
\includegraphics[width=0.8\textwidth]{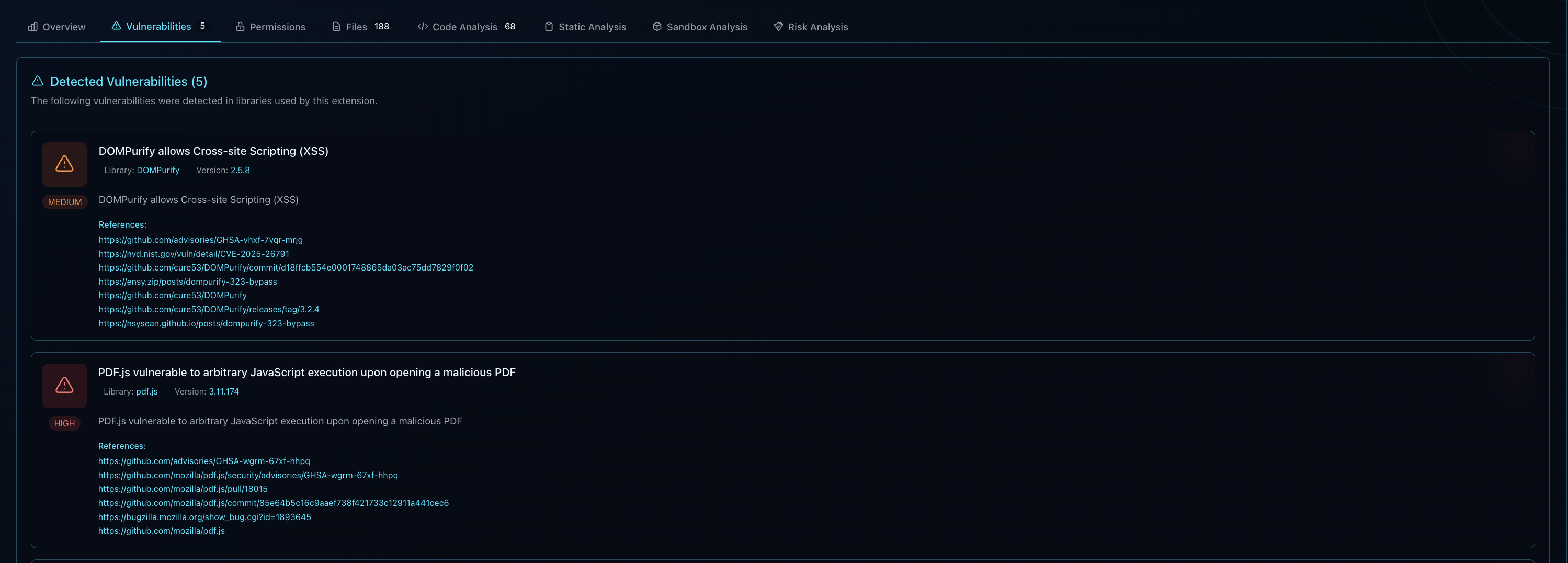}
\caption{Vulnerabilities Tab}
\label{fig:vulnerabilities_tab}
\end{figure}
\item A \textbf{Permissions Summary} (if applicable) detailing each permission required by the extension with extensive descriptions.
\begin{figure}[h]
\centering
\includegraphics[width=0.8\textwidth]{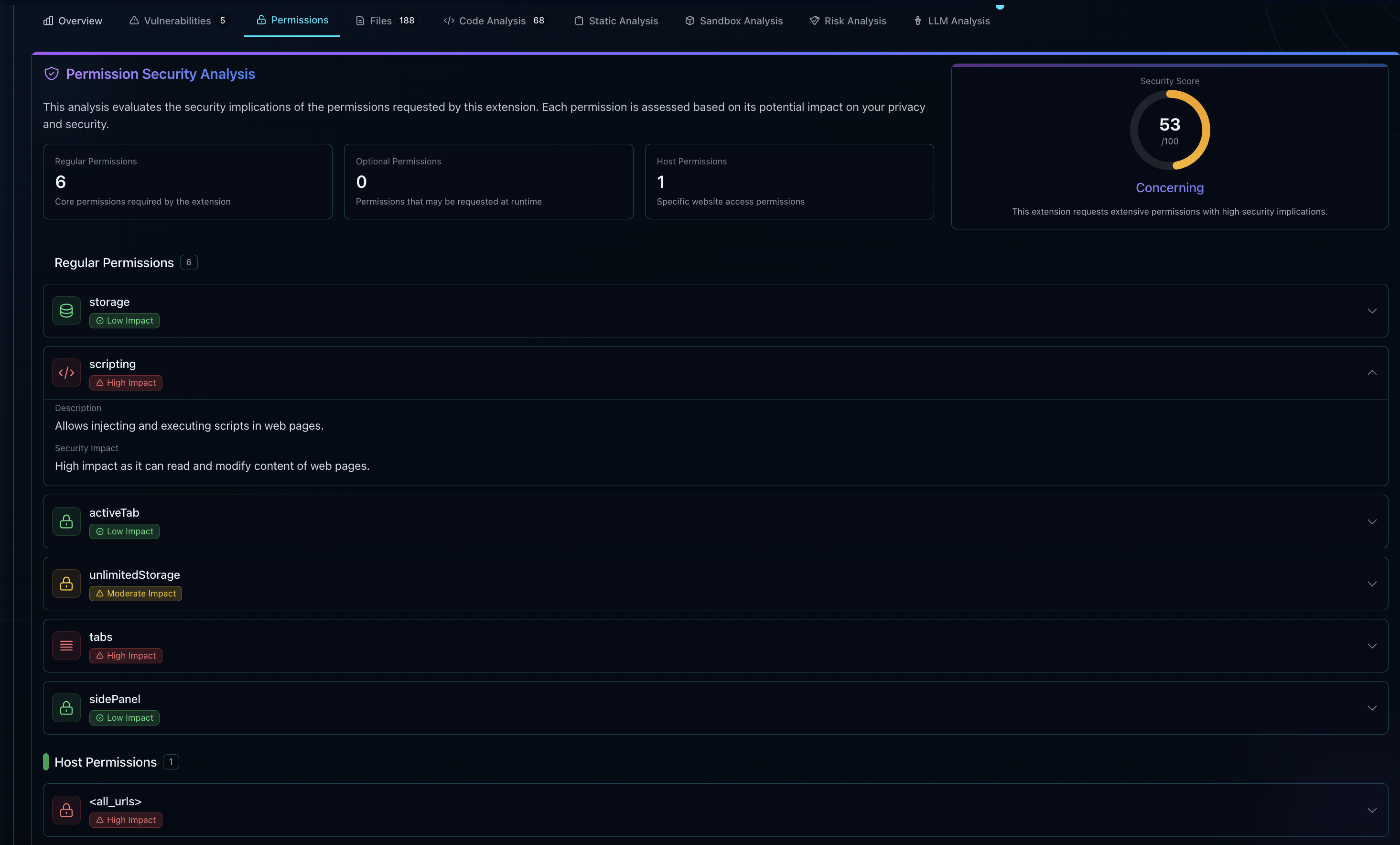}
\caption{Permissions Tab}
\label{fig:permissions_tab}
\end{figure}
\item A \textbf{File Explorer} enabling users to view all files along with their complete contents.
\begin{figure}[h]
\centering
\includegraphics[width=0.8\textwidth]{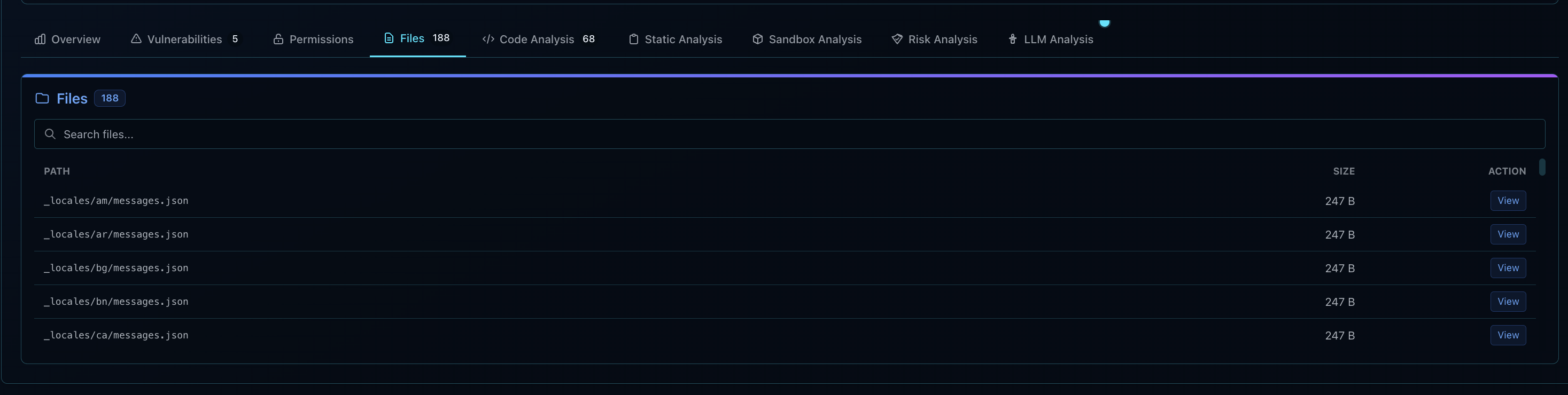}
\caption{File Explorer Tab}
\label{fig:files_tab}
\end{figure}
\item A \textbf{Code Analysis} tab displaying each file alongside detailed Abstract Syntax Tree (AST) analyses.
\begin{figure}[h]
\centering
\includegraphics[width=0.8\textwidth]{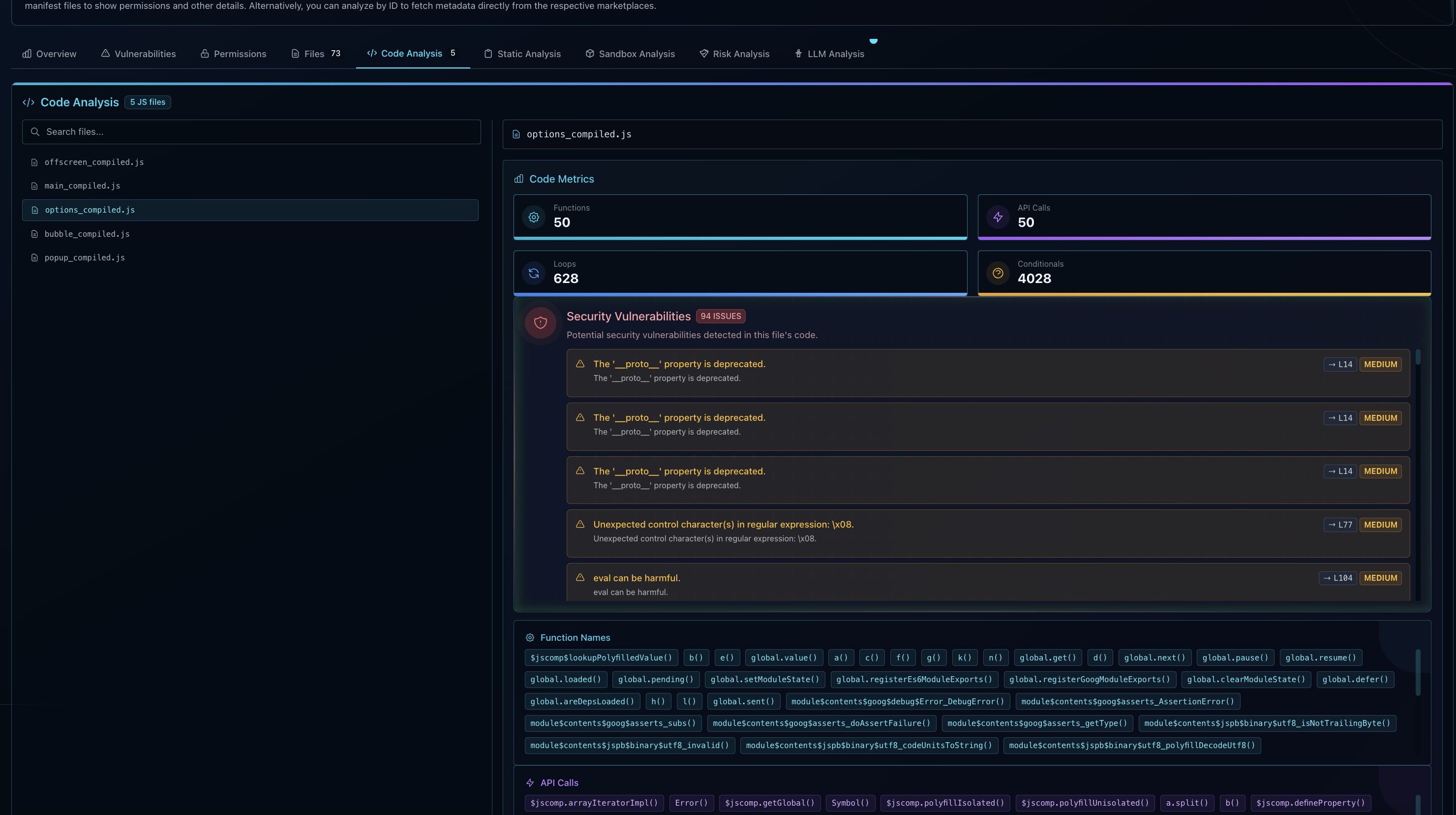}
\caption{Code Analysis Tab}
\label{fig:codeanalysis_tab}
\end{figure}
\item A \textbf{Static Analysis} tab showcasing all detected code patterns.
\begin{figure}[h]
\centering
\includegraphics[width=0.8\textwidth]{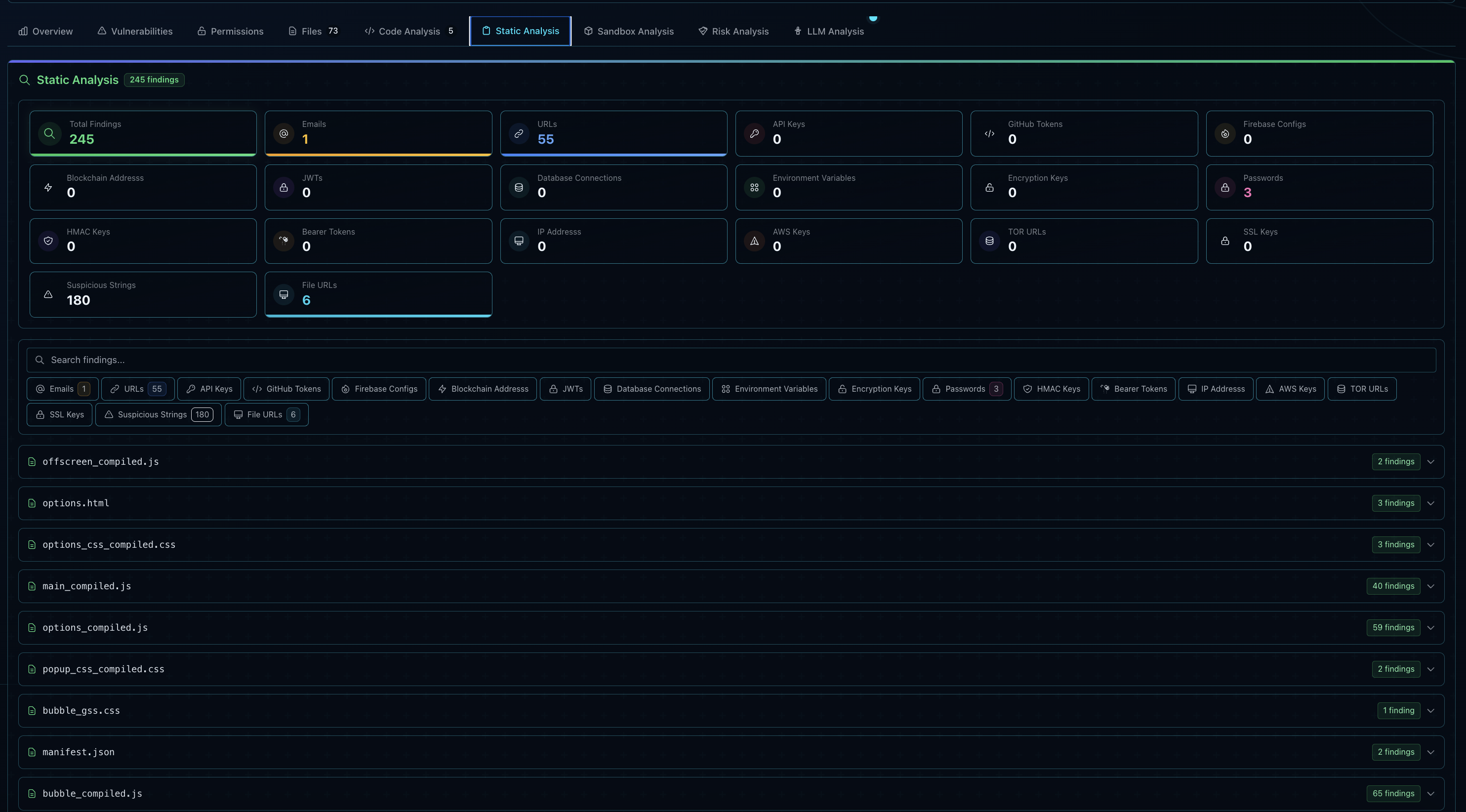}
\caption{Static Analysis Tab}
\label{fig:staticanalysis_tab}
\end{figure}
\item A \textbf{Sandbox Analysis} providing comprehensive insights derived from sandbox executions.
\begin{figure}[h]
\centering
\includegraphics[width=0.8\textwidth]{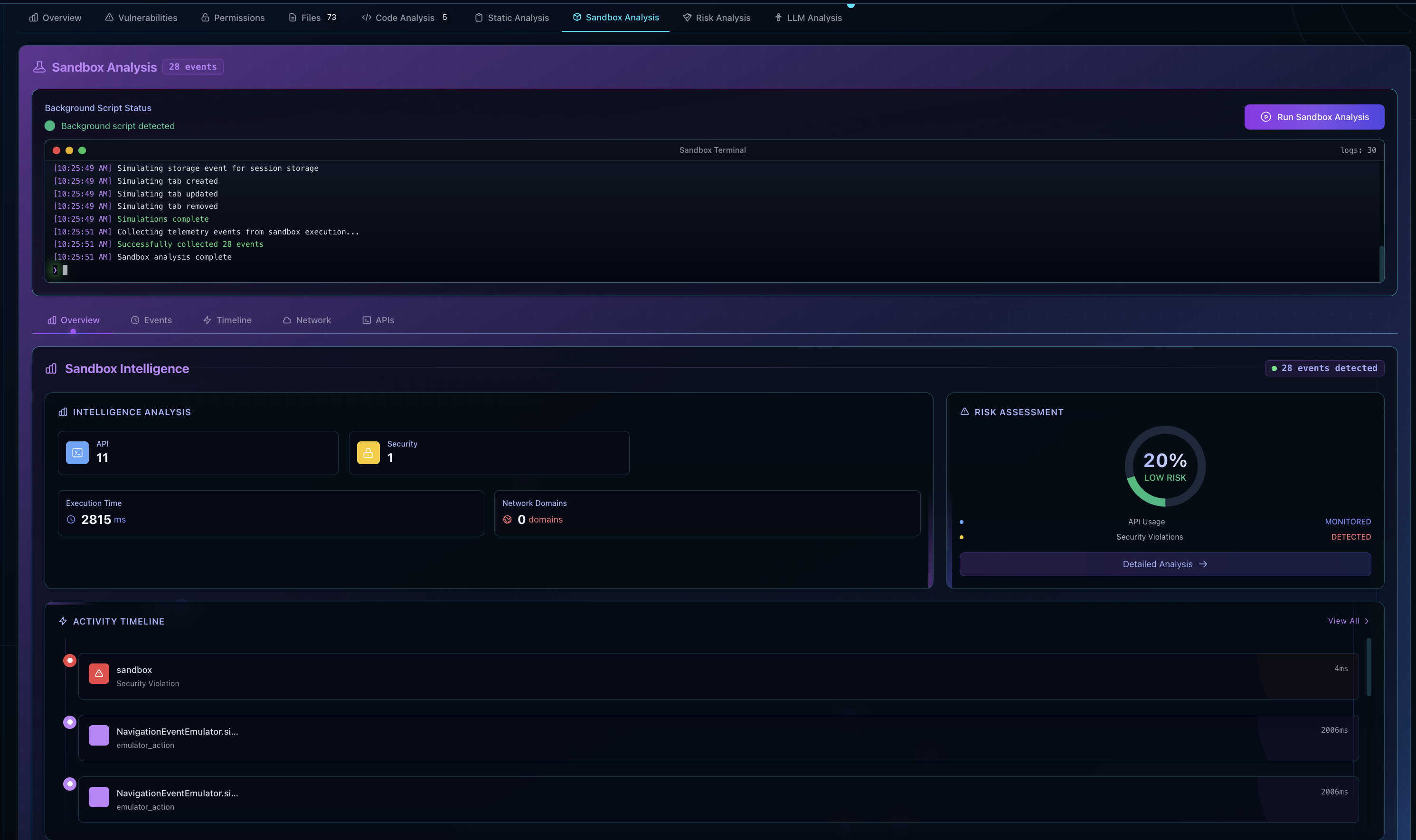}
\caption{Sandbox Analysis Tab}
\label{fig:sandbox_tab}
\end{figure}
\item An \textbf{LLM Analysis} tab evaluating both risk factors and privacy policies.
\end{itemize}
\begin{figure}[h]
\centering
\includegraphics[width=0.8\textwidth]{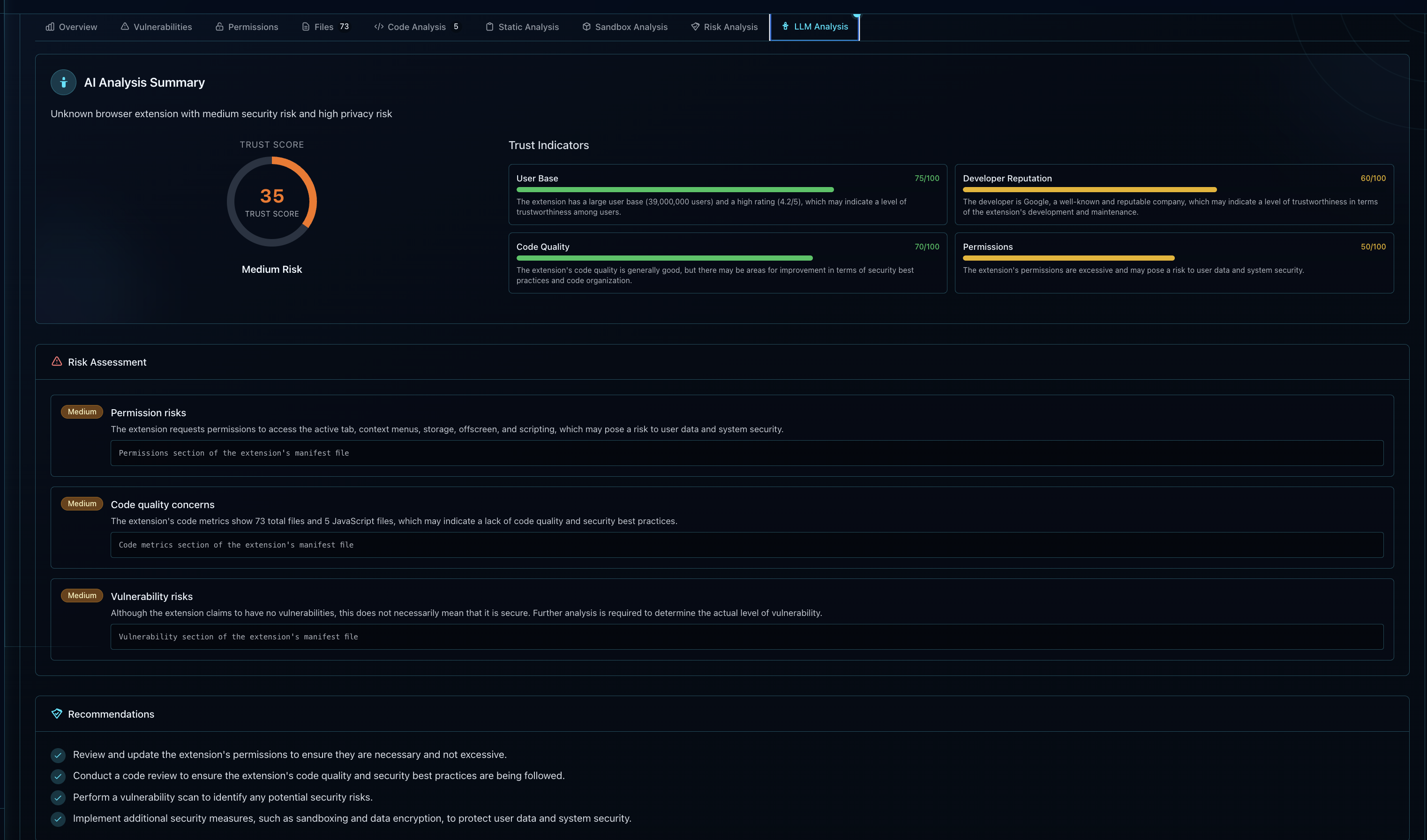}
\includegraphics[width=0.8\textwidth]{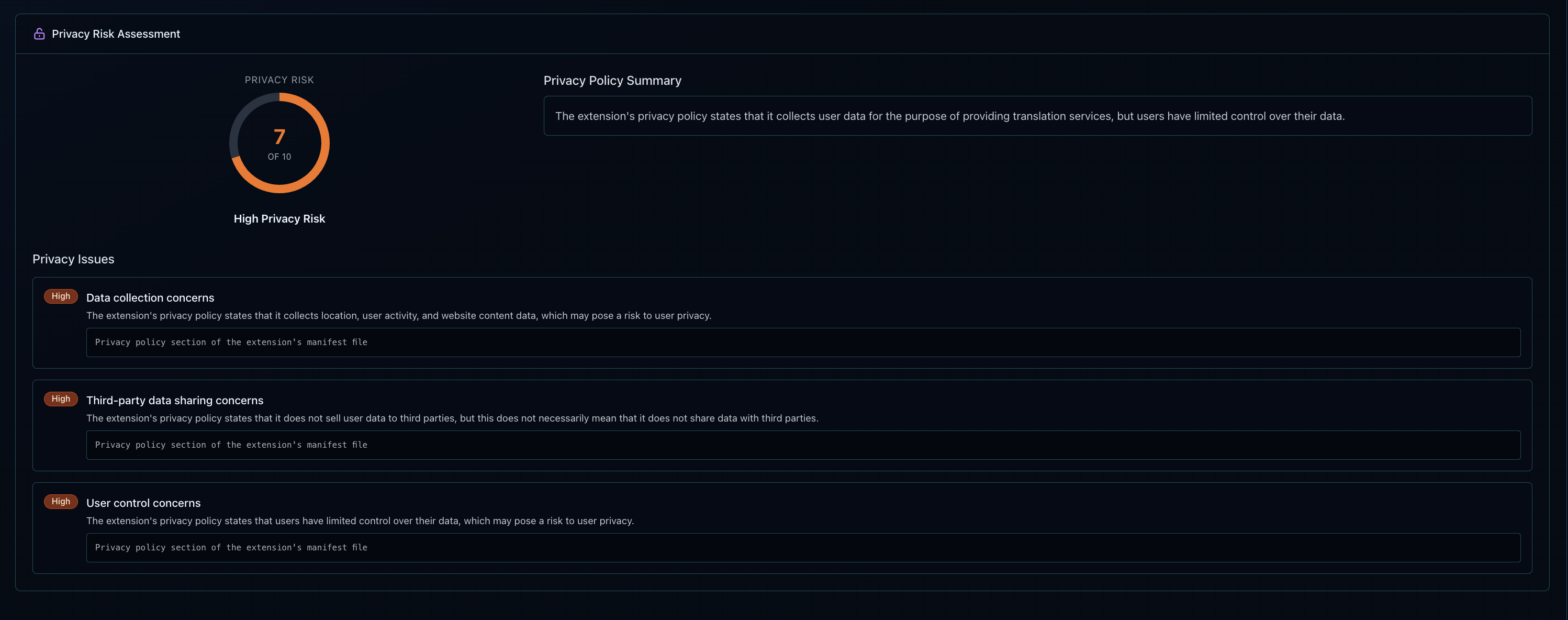}
\caption{LLM Analysis Tab}
\label{fig:llm_tab}
\end{figure}

The GUI supports interactive analysis, allowing users to toggle specific simulation parameters and re-execute analyses—for instance, simulating future dates or specific site visits (e.g., simulating a visit to \texttt{facebook.com} while the extension runs). Such interactions trigger sandbox re-evaluations under new conditions, thereby dynamically updating results. In our prototype, these interactive features enable users to explore "what-if" scenarios, examining potential behaviors of extensions under varying circumstances.

Designed primarily for developers and security analysts individuals already concerned with extension security, the interface ensures usability through clear communication. However, the LLM-generated summaries and explicit labeling of issues (e.g., \enquote{Detected data exfiltration} or \enquote{Uses high-risk API: \texttt{child\_process}}) also facilitate informed decision-making among less experienced users.

Furthermore, JavaSith is structured to support workflow integration. Potential implementations include a meta browser extension capable of scanning other extensions or a Visual Studio Code (VSCode) extension performing installation-time scans to notify users promptly. While complete workflow integration remains reserved for future research, the underlying architecture readily supports these capabilities.

\section{Technical Implementation}

JavaSith is implemented using a combination of TypeScript, JavaScript, and WebAssembly to achieve optimal performance and security. The sandbox environment leverages Secure ECMAScript (SES) \cite{js_ses}, a hardened subset of JavaScript designed to facilitate secure isolation by enforcing strict confinement and capability-based security models. Specifically, SES allows JavaSith to accurately emulate and isolate browser and Visual Studio Code (VSCode) contexts within a robust and controlled execution environment.
Key technical challenges and how we addressed them are discussed below.

\subsection{Sandboxing and API Hooking}
To intercept extension API calls, we created wrappers for Chrome’s extension APIs. For example, the extension might call \texttt{chrome.storage.local.get}; our wrapper intercepts this, logs it, possibly feeds it predefined dummy data (since in the sandbox there is no real prior storage unless we set some), and then returns a promise/result as expected. This required implementing enough of the Chrome API surface to satisfy typical extensions. We focused on APIs commonly used by malicious extensions in past cases: \texttt{chrome.storage}, \texttt{chrome.cookies}, \texttt{chrome.declarativeNetRequest} (which was used in the Nimble Capture case to strip CSP), and basic \texttt{chrome.tabs}/\texttt{chrome.runtime} messaging.

For VSCode, we created a mock \texttt{vscode} module that has stubs for the key VSCode API objects (like \texttt{window}, \texttt{workspace}, \texttt{commands}). If an extension tried to register a command or read the active text editor, we either emulate it or log the access. We allowed certain benign calls to go through (like writing to an output channel just writes to a buffer we capture).

Network requests were captured by setting a global \texttt{XMLHttpRequest} and \texttt{fetch} in the sandbox (for browser context) that intercepts the URL. In Node context, we monkey-patched \texttt{http.request} and related methods. All captured network traffic is recorded, and by default we allow it to complete (so that if an extension downloads a payload, we fetch it into the sandbox) but with options to block (to avoid possibly contacting a live malicious server if not desired—though our execution is in a isolated container, contacting the real server could alert the attacker or cause unintended effects. We often use a setting to redirect such calls to a dummy local server or record them without actually fetching).

\subsection{Time Control Mechanism}
The Time Machine required control over JavaScript’s event loop. We used Node’s ability to manipulate the event loop with libraries and a custom scheduler for timers. In practice, we intercept \texttt{setTimeout/setInterval} as shown earlier, and maintain a priority queue of scheduled tasks. If the user chooses to fast-forward time, we immediately execute all tasks in the queue. When the sandbox is running normally, we by default fast-forward any timer with delay beyond a threshold (say 1 second) to not slow analysis. However, we also incorporate a pseudo-real-time mode for observing behavior that is time-sensitive (like short intervals or animations, though those are rare in malicious code).

One tricky scenario was recursive scheduling: some malware might schedule a short timer repeatedly to create a long delay (like schedule a 1-minute timer 60 times to wait an hour). Our time acceleration can handle this by jumping in increments.

We tested the time machine on benign extensions that use alarms (Chrome’s \texttt{chrome.alarms} API) and confirmed we can trigger those immediately as well by hooking the alarm creation and triggering the callback.

\subsection{LLM Integration Details}
We integrated WebLLM by loading a model (in our test, a 8B parameter variant of LLaMA fine-tuned on code explanations) in the browser. The model runs at a few tokens per second on a modern GPU; generating a full analysis (~300 tokens) takes under a minute, which is acceptable post-analysis. Because loading such a model can be heavy, we make the LLM step optional or on-demand. In a headless mode (no GUI), JavaSith can output a textual report including the LLM's summary, and this can be skipped for quick runs.

We faced some token limit issues with the LLM context; to handle large code input, we do not feed the entire codebase. Instead, we feed summary info and only relevant code excerpts (the ones with findings or suspicious patterns). This keeps the prompt size manageable. Improving the prompt engineering and possibly fine-tuning specifically for extension analysis is future work.

\subsection{Performance Considerations}
Running full dynamic analysis on an extension can range from a few seconds to a few minutes, depending on complexity. We instrumented the sandbox to track performance overhead. For example, a simple extension that just adds a button to Gmail was analyzed in under 5 seconds, whereas an extension that loads a large library and runs multiple background tasks took ~30 seconds with time acceleration (because it scheduled many tasks). The overhead of logging and hooking was roughly a 2x slowdown on average compared to native execution—this is acceptable since we are not in a real-time scenario.

Static analysis (Retire.js scanning and regex checks) is very fast (a few seconds even for large codebases). The LLM analysis is the slowest single component (as noted, up to a minute), but this can run in parallel with user reviewing other results, or be skipped for quick runs.

We also considered the size of extensions: some have megabytes of assets (images, etc.). We ignore non-code assets aside from noting their presence. If an extension bundles minified code, our static analyzer tries deobfuscation (e.g., prettifying, or at least highlighting that it's minified/obfuscated). Extremely large code (like a big minified library) is flagged but not fully parsed to avoid slowdowns.

Memory overhead for sandboxing can be a factor if an extension is doing heavy in-memory tasks. Our sandbox by default limits memory (via container constraints) to prevent runaway processes.

\section{Case Studies of Known Attacks}
We applied JavaSith to analyze scenarios inspired by real-world attacks to illustrate its capabilities. In each case, we either reproduced the malicious code from reports (when available) or created a lightweight approximation based on descriptions in the literature, and then ran our tool to see if it would catch the malicious behavior.

\subsection{Browser Extension Supply-Chain Attack (Cyberhaven Incident)}
For this case, we obtained code snippets resembling the malicious update of the Cyberhaven Chrome extension as described by Darktrace and others. The malicious extension’s behavior included:
\begin{itemize}
    \item Exfiltrating cookies and session storage from certain high-value websites (Facebook, Google Ads, etc.).
    \item Beaconing out to an attacker-controlled domain that impersonated a legitimate service.
    \item Possibly disabling security features (one part of the campaign involved removing Content Security Policy headers).
\end{itemize}

We ran JavaSith on this code in our browser extension sandbox. The static analysis immediately flagged the presence of code accessing browser cookies and the use of an XMLHttpRequest posting data to a suspicious URL (the domain \texttt{cyberhavenext.pro} was on our known-malicious list from OSINT). The static analyzer also highlighted that the code was specifically checking if the current site was \texttt{facebook.com} (a domain trigger).

During dynamic execution, we simulated visiting \texttt{facebook.com} by triggering page navigation events in our sandbox browser context while the extension was active. JavaSith’s content script monitor captured the extension injecting a script into that page and reading document cookies. The sandbox blocked the actual network exfiltration but logged the attempt: “POST request to \texttt{https://cyberhavenext.pro/...} with payload size 5KB”. This would be a clear red flag to any analyst.

We also simulated time by setting the system date to December 25, 2024, to match the incident timing. (In this malware, time of year was not a coded trigger, it was just the attackers' choice of holiday timing. But if it had a date check, we could have adjusted accordingly.)

The LLM risk analysis for this case produced a summary: \textit{“This Chrome extension appears to steal authentication tokens (cookies) from the user’s browsing sessions on specific sites and send them to an external server under the guise of a legitimate domain. This behavior indicates a serious account takeover risk. The extension operates silently and could grant attackers unauthorized access to user accounts. This is highly malicious.”} This aligns with the assessment reported by Nightfall’s post-mortem.

In a real deployment, JavaSith would have caught this malicious update as soon as it was analyzed, potentially preventing the days of undetected operation. One can imagine a company using JavaSith to automatically scan updates to extensions before allowing them in a managed browser environment, which could have flagged the Cyberhaven extension update as suspicious due to its new network behavior (especially since previous versions likely did not have such behavior).

\subsection{Malicious VS Code Extensions (Cryptomining Campaign)}
We tested JavaSith on a sample malicious VSCode extension patterned after the April 2025 cryptominer case. The extension’s \texttt{extension.js} on activation would:
\begin{enumerate}
    \item Download a PowerShell script from \texttt{asdf11.xyz}.
    \item Execute it by spawning \texttt{powershell.exe} via Node’s \texttt{child\_process.exec}.
    \item After the script finishes, call VSCode’s commands API to install another extension (the real one it was faking).
\end{enumerate}

JavaSith’s static analysis flagged multiple issues right away: a hardcoded URL containing an uncommon domain (likely malicious), usage of \texttt{child\_process.exec} to run PowerShell, and the presence of a base64-encoded string which turned out to be part of the script. These by themselves put the risk at High.

In the sandbox, when executed, the extension indeed invoked the PowerShell. Our isolated environment prevented launching a real PowerShell process, but we intercepted the command string. It was a complex script; JavaSith logged attempts to modify registry keys and schedule tasks (the script content was decoded from base64 at runtime; our dynamic instrumentation caught the decoding function and captured the decoded result).

One interesting event was the extension calling a VSCode command to install an extension by ID. The sandbox’s fake VSCode API recorded this call Command: 
\begin{lstlisting}
workbench.extensions.installExtension
\end{lstlisting}
with argument \textit{ms-vscode.prettier}”). This is an unusual action for an extension to perform, so our heuristic flagged it as well (“Extension installing another extension”).

The Time Machine wasn’t heavily utilized here because the malware ran quickly on activation. However, had the cryptominer delayed its start (say, only run after VSCode has been open for 10 minutes to avoid immediate suspicion), our accelerated timers would trigger it.

The LLM summary for this case was along the lines of: \textit{“This VSCode extension executes a hidden PowerShell script that makes system changes (disabling updates, adding tasks) and then downloads a cryptomining program. It also installs the legitimate Prettier extension afterward, likely to hide its malicious actions.It seems like this is a malicious extension that compromises the host machine for cryptojacking.”}

This matches the description from Symantec/Broadcom. JavaSith successfully identified the malicious behavior, and an end-user running it would be alerted before the miner could actually run (especially since our sandbox blocked the actual external network calls and process launches by default after logging them).

\subsection{Malicious VS Code Extensions (Information Stealers)}
We next took a scenario of the info-stealing extensions. Using a simplified reimplementation of `clipboard-helper-vscode` (as per ReversingLabs description), we created an extension that registers an event for text copy (simulated via a periodic clipboard check) and, if it sees any text, sends it via an HTTP POST to a Discord webhook URL.

JavaSith’s static scan immediately spotted the Discord webhook URL string embedded in the code. This triggered one of our high-severity rules. During dynamic run, we put some text in the simulated clipboard and triggered the extension’s function. The sandbox logged an HTTP POST to \texttt{discord.com/api/webhooks/...} with a payload containing the test text. We prevented the actual network call for safety.

The outcome was straightforward: JavaSith categorized this extension as malicious with the reason “exfiltrating user data to an external server (Discord webhook)”. The LLM summarized: \textit{“This extension appears to send clipboard data to a remote server (using a Discord webhook) without permission. This behavior is indicative of spyware, potentially stealing sensitive information. Users’ copied data (which may include passwords or tokens) could be compromised.”}

Because the extension code was otherwise not doing anything useful (in our test), it was clearly malicious. In real cases, such extensions might also provide some fake functionality to look legitimate. JavaSith would still catch the background stealing unless it was extremely stealthy or heavily encrypted (in which case static patterns might still detect something anomalous).

\subsection{Malicious NPM Package}
Finally, we tested a malicious NPM scenario inspired by the `os-info-checker-es6` package. We constructed a package that on install checks the OS, then downloads a payload if on Windows. We introduced obfuscation by using zero-width spaces in the code (to mimic the Unicode hiding trick). We also hosted a fake “payload URL” via an internal test server.

JavaSith’s static analysis caught unusual Unicode characters in the source (non-printable characters, which we flag as obfuscation). It also saw that the install script had a `curl` command (in our simulation) to a Google Calendar URL (mimicking the technique of using Google services as a C2). These were flagged as suspicious.

During dynamic install in the sandbox, the package tried to perform the HTTP GET for the payload. We intercepted that and responded with a dummy response to allow it to continue. The package then wrote a file to the temp directory (which in a real attack could be a malicious binary) and attempted to execute it. Our sandbox prevented execution, but we captured the attempt (“exec file: ./payload.exe”).

The Time Machine wasn’t needed here, as the package ran its malicious logic immediately. JavaSith effectively identified it as malicious. Notably, no normal package should hide code in such a manner or reach out to an unrelated external URL in an install script. Our tool would warn the user not to install this package.

These case studies collectively demonstrate that JavaSith can handle a range of malicious behavior: from web-focused data theft to host compromise through script execution, across different environments.

\section{Evaluation}
We evaluate JavaSith on two fronts: detection effectiveness and performance overhead. Our evaluation is limited by the availability of known ground-truth malicious extensions and packages (often such code is not public). However, using samples from threat reports and synthetic variants, we can gauge how well JavaSith performs.

\subsection{Detection Coverage and Effectiveness}
We compiled a test set of 20 extensions/packages:
\begin{itemize}
    \item 5 known malicious Chrome extensions (from public sources or recreated from descriptions).
    \item 5 benign Chrome extensions (popular ones like uBlock Origin, etc.) as false-positive tests.
    \item 3 malicious VSCode extensions (including cryptominer and info-stealer types).
    \item 3 benign VSCode extensions (e.g., Python extension, Prettier).
    \item 4 NPM packages (2 malicious or suspected, 2 benign).
\end{itemize}

JavaSith successfully flagged all malicious samples with high severity. In cases where we had actual known malicious code, the dynamic analysis often produced very evident logs (e.g., network calls to suspicious domains, attempts to execute shell commands). The static analysis was particularly useful for those that attempt to hide until certain triggers. For example, one malicious Chrome extension in our set was known to only activate on specific domains; our static scan found those domain strings in the code.

We did not encounter any false negatives in this set, though we acknowledge the set is not exhaustive. The benign extensions, in contrast, yielded no high-severity alerts. There were a few low-severity notices (for instance, one benign extension included Google Analytics code which made network requests; our tool logged it but our heuristics recognized the domain as benign and the behavior as expected). The LLM sometimes was slightly over-cautious; for a benign extension that accessed a web API, the LLM mused if it might leak data, but our final risk scoring logic did not mark it as malicious because none of the concrete detectors fired. This highlights the importance of combining automated flags with the LLM reasoning, but not relying solely on the latter.

A metric we consider is whether JavaSith would have caught known incidents prior to their public discovery:
- The Cyberhaven extension: Yes, likely, due to cookie access and strange network traffic.
- The VSCode miner extensions: Yes, the combination of static (PowerShell present in code) and dynamic (network and process spawn) would flag it.
- The Discord webhook info-stealers: Absolutely, given the static presence of the webhook URL.
- Various NPM attacks (like the Unicode one, or packages that steal tokens): Yes, as long as we run the install, their activity becomes evident.

One limitation is that if an extension’s payload is extremely environment-specific (for example, only triggers if a certain API returns a specific response), we might not trigger it in analysis. However, our multi-pronged approach (especially the static code reading via LLM) can still notice something’s off.

\subsection{Performance Overhead and User Experience}
We measured the analysis time for a subset of extensions under different settings (with/without time acceleration, with/without LLM):
- For small extensions (< 100KB code), JavaSith completes analysis in 5--10 seconds without LLM. With LLM, add ~30 seconds for model loading and inference.
- For larger extensions (~1MB of code/assets), dynamic analysis might take ~20 seconds, static ~5 seconds, plus LLM ~30--60 seconds. So under 2 minutes end-to-end.

While this is slower than a trivial static linter, it is quite reasonable given the depth of analysis. In a scenario where a user is installing a new extension, a 1-2 minute scan is a minor inconvenience compared to the potential damage of a malicious install.

We also considered the sandbox overhead. Running a cryptominer in sandbox, for example—our container prevented actual CPU mining, but we let the process spawn to observe behavior for a short time then terminated it. The overhead on the host system was negligible beyond the work done inside the container (which we limited). The user’s system might see some CPU use during analysis (especially if the LLM runs on CPU, though WebGPU can offload to GPU if available).

The GUI’s responsiveness was tested with logs of ~1000 events (which was our biggest case—the cryptominer extension performed many actions). The interface remained navigable after we optimized how we render logs (batching and lazy-loading entries).

Memory consumption: the largest factor is the LLM model (which can be several gigabytes loaded). For everyday use, users might opt not to load the LLM unless they specifically want the detailed analysis; even without it, the rest of the tool functions. We also foresee that smaller models or quantization will reduce this footprint (and as hardware improves, running these models will become more feasible; on a high-end machine a 7B model runs sufficiently).

\subsection{Limits of Detection}
No security solution is perfect. We identified a few scenarios where JavaSith might not fully catch the threat:
- If a malicious extension is completely dormant until it receives a remote command (e.g., it polls a server for a signal to activate), and if that command is not provided during analysis, we might only see the polling (which itself could be deemed suspicious, but if it's a common benign behavior like checking for update, it could be overlooked). One mitigation is to examine the code for any hints of such logic and allow analysts to feed custom inputs (like simulating a command from a C2 server).
- Highly obfuscated code could stymie static analysis and even the LLM. We had one test with heavily obfuscated code (using a packer); the static analysis basically flagged “code is heavily obfuscated” and the LLM couldn’t parse it meaningfully. In such cases, dynamic analysis is even more crucial, but if the code is obfuscated to the point it constructs payload at runtime in a convoluted way, only the raw behavior will show. JavaSith can still log that behavior but understanding it is hard. In future, integrating a de-obfuscator or using the LLM to attempt de-obfuscation could help.
- Resource-intensive or interactive extensions (like those requiring user UI interaction to trigger functionality) might not reveal much in an automated run. For instance, an extension that only does something when a browser action button is clicked. Our current sandbox doesn’t simulate clicking the extension’s button. We rely on static analysis in such cases, or a user manually invoking some action. Extending the sandbox to simulate common user interactions (like triggering commands or clicking an extension UI if present) is potential future work.
- Native code: If an extension or package includes a compiled binary (e.g., a Node native addon or a WASM module), our analysis of its internal logic is limited. We can detect its presence and maybe intercept calls it makes (if through known interfaces), but analyzing compiled payloads is out of scope. At least, we flag that such a binary exists, which is itself unusual for many extensions.
- Evasion: A sophisticated attacker might try to detect analysis environments. For example, checking if certain debugging objects exist or if known sandbox artifacts are present. We attempted to make the sandbox as transparent as possible, but certain differences (like timing or the lack of real user data) could theoretically tip off the malware that it’s under analysis. This cat-and-mouse game is common in malware. Since JavaSith runs on the client side without a standard fingerprint, attackers would have a very hard time targeting it.

Overall, our evaluation indicates JavaSith is effective in identifying malicious behavior in extensions and packages, with manageable runtime overhead. In the next section, we discuss the broader context, limitations, and how JavaSith could evolve.

\section{Discussion}
JavaSith represents a step toward giving end-users and enterprise security teams more control and visibility into the extensions and packages they use. By performing client-side analysis, we avoid reliance solely on marketplace gatekeepers. Here we discuss the implications of such a tool and some broader considerations.

\textbf{Empowering Users vs. Usability:} One might question if average users would run a 1-minute analysis before installing a browser extension. Power users or security-conscious organizations might, but regular users probably will not. However, JavaSith could be integrated into browsers or IDEs as an automated vetting system. For example, enterprise browser deployments could automatically run JavaSith on any extension before allowing it. This way, the burden is not on the end-user to manually invoke it each time; it becomes an invisible shield that only alerts when something is wrong. Similarly, a developer’s IDE like VSCode could integrate such scanning when new extensions are installed (perhaps in a future ``Secure Mode'' of VSCode, analogous to Workspace Trust).

\textbf{False Positives and Trust Calibration:} A challenge in any detection system is avoiding false alarms. We tuned JavaSith to prioritize catching true malicious intent, which means it may raise warnings for things that are not necessarily malicious but “could be.” For example, an extension that reads cookies might be doing so for legitimate reasons (like an extension that manages session cookies). JavaSith would flag that, but our LLM or explanation can provide context (“Extension reads cookies for domain X – this could be normal if functionality requires it”). We expect that over time, with a larger corpus, we can refine heuristics to better differentiate benign vs malicious patterns (possibly even training the LLM on known-bad vs known-good code to see differences).

We also allow a “learning mode” where if a user knows an extension is safe, they can mark it, and our system remembers that decision (similar to how antivirus allows exceptions). However, this can be risky if users whitelist something malicious, so it’s more viable in a managed setting where security teams review and approve certain extensions.

\textbf{Privacy and Offline Use:} A key reason to have a client-side LLM is privacy (and independence from cloud). JavaSith never needs to send the extension’s code to an external server for analysis, which is important for sensitive environments (e.g., evaluating an extension in a corporate setting where the code might have proprietary info or where contacting an external API could leak that someone is investigating a particular extension). The flip side is performance constraints of local analysis, but as hardware improves and specialized security models are developed, this will get better.

\textbf{Keeping Pace with Attackers:} Attackers will undoubtedly innovate. JavaSith should be updated with new detection logic as new techniques emerge (similar to antivirus updates). For instance, if attackers start using more exotic covert channels (like using browser extension storage in a clever way), we would integrate checks for that. Our modular architecture (especially the static analysis rules and LLM prompts) can be updated without overhauling the whole system. The LLM component could even be used in a threat-hunting mode where it’s asked “do you see anything potentially malicious or unusual in this code?” to catch novel techniques that don't match explicit rules.

\textbf{Integration with Threat Intelligence:} We could enhance JavaSith by feeding it known bad indicators (domains, hashes, etc.) from threat intel feeds. For example, if a certain URL or regex is known to appear in malware, our static scanner can have that in a blacklist. Similarly, if a certain obfuscation signature (like a known packer stub) is detected, we can label it. The dynamic part could also output traces in a format that can be cross-checked against threat databases.

\textbf{User Interface and Actionability:} One risk of presenting too much information is overwhelming the user. We aim to have clear, concise risk outputs. If JavaSith finds nothing major, it could simply say “No significant risks detected.” If it does find serious issues, it should summarize: “Likely Malicious – reasons: contacts known bad site, steals data.” For an average user, that alone might prompt them to avoid the extension. For a developer or researcher, the detailed logs remain available. In testing, we found the combination of a risk score (High/Medium/Low) plus a one-sentence reason derived from LLM output is effective.

\textbf{Limitations of LLM Analysis:} While powerful, the LLM is not infallible. It might hallucinate interpretations that aren't accurate. We mitigate this by structuring the prompt to stick to evidence (we list the events and code facts for the model). We also allow the user to see exactly what events were fed into the LLM, for transparency. If the LLM says “steals Facebook token” and the user sees indeed a network call to a suspicious domain after accessing Facebook, that aligns. If the LLM says something the user doesn’t find in logs, that could indicate a misinterpretation, which is why raw data is always available.

Interestingly, the LLM can sometimes suggest remediation (like “you should remove this extension immediately”), which is helpful. It might also suggest other steps (revoke credentials, etc., if an extension was found stealing tokens). These touches turn a raw analysis tool into more of an assistant.

\textbf{Open Sourcing and Community:} A tool like JavaSith would benefit from community contributions. If open-sourced, security researchers could contribute new rules (like volunteer-maintained lists of malicious patterns). Also, being transparent about how it works can increase trust. Attackers might read the code and attempt to circumvent it, but since they already assume they may be under scrutiny, the benefit of community-driven improvements likely outweighs the risk. It's a constant arms race anyway, and openness can speed defensive innovation.

\textbf{Broader Platform Coverage:} While we focused on Chrome and VSCode (and indirectly Node/NPM), the general approach can extend to other ecosystems:  other IDE plugins and other package managers (PyPI, RubyGems). Given more time, we plan to adapt our sandbox to  scan mobile app plugins or browser user scripts. The supply chain threat is everywhere, and each environment has its specifics but similar core issues (unvetted code running with user privileges).

\textbf{Continuous Monitoring vs Pre-Install Scan:} JavaSith currently performs a pre-install or on-demand scan. Another approach is continuous monitoring of extension behavior at runtime on the user's system. While conceptually similar, that veers into intrusion detection rather than preemption. There is value in both; a sophisticated enterprise might deploy JavaSith in CI pipelines (to scan dependencies) and on endpoints (to watch extension behavior live). However, running the sandbox continuously is expensive. Instead, a lighter-weight runtime monitor (like an extension that monitors other extensions, or OS-level monitors for processes launched by VSCode extensions) could complement JavaSith’s heavy upfront scan.

In conclusion, JavaSith opens new possibilities for proactive defense in the extension and package ecosystem. It is not a silver bullet, but it raises the effort required for attackers to succeed undetected. As we refine the tool and gather more feedback, we expect it to become a practical addition to the security toolkit of developers and IT administrators alike.

\section{Limitations and Future Work}
While our work demonstrates the feasibility and value of client-side extension analysis, there are limitations to acknowledge and areas for improvement:

\textbf{Scalability:} JavaSith currently analyzes one extension or package at a time. In an enterprise with hundreds of extensions in use, scanning them all periodically could be heavy (though it can be parallelized across machines or time). We plan to add a batch mode and perhaps a “quick scan” mode for routine re-checks (e.g., skip dynamic analysis if the code is unchanged from a previously scanned version, etc.).

\textbf{Depth of Simulation:} Our browser simulation is not a full Chrome browser – some extension behaviors (like complex interactions with actual page content or advanced Chrome APIs) might not be fully covered. We intend to integrate more of Chrome’s own headless capabilities; possibly even using Chrome’s headless mode with the extension loaded and then instrumenting it via the DevTools protocol (which some research tools have done). Similarly for VSCode, while we simulate enough for typical malicious patterns, an extension doing heavy GUI work in VSCode might not see all its functionality in our sandbox (though that’s less relevant to detecting malicious activity).

\textbf{Evasive Malware:} If an extension uses extremely subtle triggers or requires specific user context, our automated analysis might miss it. One idea is to incorporate feedback-guided execution: have JavaSith try multiple scenarios (e.g., simulate browsing a list of popular sites, simulate user inactivity or specific keystrokes) to coax out hidden behavior. This is an area of future research, essentially fuzzing the extension's event space.

\textbf{Machine Learning Enhancements:} The current LLM integration uses a general-purpose model. Training or fine-tuning an LLM specifically on malicious vs. benign extension code (if we gather such a dataset) could improve its accuracy and reduce false positives/negatives. We also consider adding a smaller, rule-based ML classifier that runs before the LLM to handle obvious cases faster.

\textbf{User Study and Feedback:} We plan to conduct user studies with developers and security analysts to gauge JavaSith’s usability. Understanding how users interpret the risk reports, and whether it actually influences their decisions (e.g., not installing an extension), is key. If users find parts of the output confusing, we will refine the presentation.

\textbf{Integration with Dev Tools:} We envision future integration such as a browser extension that automatically runs JavaSith when you try to install another extension from the web store, or a VSCode feature that scans extensions in the background. Collaborating with browser/IDE vendors to embed such capabilities (even in a lite form) would greatly increase impact. For now, JavaSith remains an external tool, but one that could be automated via scripts.

\textbf{Extending to Other Platforms:} As noted, other extension systems could be targeted by similar supply chain attacks. In future work, we aim to support at least Mobile app plugins or mods (like browser mods on Android) are another frontier.

\textbf{Live Protection:} A natural extension is combining JavaSith’s analysis with live protection. For example, an enterprise could run JavaSith on all allowed extensions and then deploy a policy that if an extension tries at runtime to do something outside what was observed (like contacting a new domain), it gets blocked. This kind of whitelisting could be powerful but also complex to maintain. Nonetheless, it’s an interesting direction where JavaSith's output (the expected behavior of an extension) forms a baseline for monitoring in production.

\section{Conclusion}
We presented JavaSith, a state of the art client-side framework that brings advanced analysis techniques to bear on the problem of malicious browser extensions, VSCode extensions, and NPM packages. By emulating the runtime environment and accelerating time, JavaSith can coerce malicious code into revealing itself, while static analysis and language model reasoning provide a comprehensive understanding of potential threats. Our case studies of real attacks illustrate that JavaSith would have been effective at detecting stealthy malicious behaviors that previously went unnoticed, thus potentially preventing significant breaches.

JavaSith complements existing security measures by acting as a last line of defense at the point of installation or update. As supply chain attacks continue to rise and evolve, tools like JavaSith empower users and organizations to regain visibility and control. While not a panacea, the multi-faceted approach of JavaSith raises the bar for attackers – they must now evade not just store policies but also dynamic sandboxes, static detectors, and AI-based scrutiny on the client side.

Future work will aim to broaden JavaSith’s coverage to more platforms, refine its detection logic via larger datasets and user feedback, and optimize its performance to make it practically invisible in user workflows. We also plan to explore collaborations with platform vendors to integrate aspects of JavaSith into extension marketplaces and developer tools (for example, a ``scan before install'' feature).

In summary, the JavaSith framework demonstrates that powerful analysis of extension code can be done locally, preserving user privacy while uncovering malicious intent. We hope this work spurs further development of user-centric security tools in the software supply chain domain. By shining a light into the black box of extension behavior, we can make it significantly harder for attackers to hide in plain sight.

\end{document}